\documentclass[conference]{IEEEtran}
\IEEEoverridecommandlockouts
\usepackage{cite}

\usepackage{amsmath,amssymb,amsfonts}
\usepackage{algorithmic}
\usepackage{graphicx}
\usepackage{textcomp}
\usepackage{xcolor}
\def\BibTeX{{\rm B\kern-.05em{\sc i\kern-.025em b}\kern-.08em
    T\kern-.1667em\lower.7ex\hbox{E}\kern-.125emX}}
\begin{document}
\title{Precise Insulin Delivery for Artificial Pancreas:\\ 
A Reinforcement Learning Optimized Adaptive Fuzzy Control Approach}
\author{
\IEEEauthorblockN{MAMECHE Omar, ABEDOU Abdelhadi, MEZAACHE Taqwa, TADJINE Mohamed}
\IEEEauthorblockA{Process Control Laboratory, Control Engineering Department, \\
Ecole Nationale Polytechnique, Algiers, Algeria \\
omar.mameche@g.enp.edu.dz}
\IEEEauthorblockA{abdelhadi.abedou@g.enp.edu.dz}
\IEEEauthorblockA{taqwa.mezaache@g.enp.edu.dz}
\IEEEauthorblockA{mohamed.tadjine@g.enp.edu.dz}
}
\maketitle
\begin{abstract}

This paper explores the application of reinforcement learning to optimize the parameters of a Type-1 Takagi-Sugeno fuzzy controller, designed to operate as an artificial pancreas for Type 1 diabetes. The primary challenge in diabetes management is the dynamic nature of blood glucose levels, which are influenced by several factors such as meal intake and timing. Traditional controllers often struggle to adapt to these changes, leading to suboptimal insulin administration. To address this issue, we employ a reinforcement learning agent tasked with adjusting 27 parameters of the Takagi-Sugeno fuzzy controller at each time step, ensuring real-time adaptability. The study's findings demonstrate that this approach significantly enhances the robustness of the controller against variations in meal size and timing, while also stabilizing glucose levels with minimal exogenous insulin. This adaptive method holds promise for improving the quality of life and health outcomes for individuals with Type 1 diabetes by providing a more responsive and precise management tool. Simulation results are given to highlight the effectiveness of the proposed approach

\end{abstract}


\section{Introduction}
Type 1 diabetes, characterized by the pancreas' failure to produce insulin, necessitates meticulous blood glucose management through frequent checks and insulin administration\cite{b9}. Recent advancements have introduced the artificial pancreas, a system integrating a continuous glucose monitor, an insulin infusion pump, and a control algorithm to autonomously regulate glucose levels\cite{b10}.
This innovative technology aims to alleviate the burden on individuals by automating insulin delivery, enhancing physiological and psychosocial outcomes\cite{b10}. Our research concentrates on perfecting this algorithm, which serves as the brain behind the artificial pancreas.

Current research predominantly focuses on various control strategies to enhance the accuracy, safety, and robustness of these systems. Notably, PID controllers have been extensively studied for their potential in glucose regulation. For instance, a study on the design and evaluation of a robust PID controller for a fully implantable artificial pancreas highlights the viability of PID controllers in managing glucose levels effectively under a controlled environment \cite{b1}. Further research aimed at enhancing the safety and robustness of these systems through improved PID-based control strategies underscores ongoing efforts to mitigate the risks associated with automated insulin delivery \cite{b2}.

Moreover, Model Predictive Control (MPC) has emerged as a promising alternative, with studies demonstrating its adaptability to individual patient profiles and its efficacy in compensating for meal-induced glucose fluctuations (\cite{b3} and \cite{b4}). These advancements highlight the customizability and forward-thinking approach of MPC in managing complex physiological variables\cite{b24}.

Parallel to these developments, the use of Takagi-Sugeno fuzzy logic control has been explored, reflecting a shift towards more adaptive and nuanced control mechanisms. The research titled \cite{b5} illustrates the application of fuzzy logic principles to capture the nonlinearities and uncertainties inherent in human physiology, which are often challenging for traditional models.

Furthermore, machine learning is being introduced in control engineering through new approaches such as state estimation\cite{b27}, system identification\cite{b28}, and control\cite{b29}. The integration of genetic algorithms\cite{b6} and of reinforcement learning\cite{b30} for optimizing fuzzy logic controllers illustrates the convergence of machine learning techniques with control systems, enhancing the precision and efficiency of controllers in handling dynamic and unpredictable scenarios.

This study builds upon these foundational works by introducing reinforcement learning to optimize the parameters of a Takagi-Sugeno fuzzy controller in real-time. By leveraging reinforcement learning, this research aims to surpass the adaptability limitations of traditional and current adaptive  and optimization control strategies, offering a novel contribution to the field of artificial pancreas systems.

\section{Materials and Methods}

\noindent For this article, our focus is on developing controllers to regulate glucose levels in the blood by injecting insulin. 

\noindent For this article, our focus is on developing controllers to automatically regulate glucose levels in the blood by injecting insulin. We used reinforcement learning and fuzzy logic control tools for this purpose. A brief introduction is presented below:


\subsection{Fuzzy Logic Control}

\noindent Fuzzy Logic Control (FLC) is a type of control system that employs the principles of fuzzy logic to mimic human decision-making processes. Unlike traditional binary (on/off) logic, fuzzy logic allows for a more flexible and nuanced approach to handling uncertainty and imprecision in a system\cite{b26}.

\noindent  In traditional control systems, precise mathematical models are used to represent the relationships between inputs and outputs. However, many real-world situations involve variables that are not easily defined or precisely measured. Fuzzy logic controllers are particularly useful in situations where traditional control methods may struggle due to complex models, the presence of uncertainty, imprecision, or incomplete information. This is why we opted for fuzzy logic control.

\noindent FLC has been successfully applied in various fields, including industrial control systems, consumer electronics, automotive systems, biomedical applications, and more.

\vspace{0.15cm}

The fuzzy set is a powerful tool and allows us to represent objects or members in a vague or ambiguous way \cite{b11}. The fuzzy set also provides a way that is similar to a human being’s concepts and thought process. However, just the fuzzy set itself cannot lead to any useful and practical products until the fuzzy inference process is applied. To implement fuzzy inference to a real product or to solve an actual problem, three consecutive steps are needed, which are: \textbf{Fuzzification}, \textbf{Fuzzy Inference} and \textbf{Defuzzification}. 

\paragraph{Fuzzification}

\hspace{0.01cm} \newline
 Fuzzification is the first step to apply a fuzzy inference system. Most variables existing in the real world are crisp or classical variables. One needs to convert those crisp variables to fuzzy variables, and then apply fuzzy inference to process those data to obtain the desired output. 

 \noindent Generally, fuzzification involves two processes: derive the membership functions for input and output variables and represent them with linguistic variables. This process is equivalent to converting or mapping classical set to fuzzy set to varying degrees.

 \noindent  In practice, membership functions can have multiple different types, such as the triangular waveform, trapezoidal waveform, Gaussian waveform, bell-shaped waveform, sigmoidal waveform and S-curve waveform. The exact type depends on the actual applications\cite{b12}.

\paragraph{Fuzzy Inference / Fuzzy Rule base}

\hspace{0.01cm} \newline
A fuzzy control rule can be considered the knowledge of an expert. This fuzzy rule is represented in the form of IF-THEN sequences, which describe what actions should be taken based on the currently observed information.

\vspace{0.1cm}

\hspace{1.1cm} IF (Antecedent) THEN (Consequence)

\vspace{0.1cm}

\noindent A fuzzy IF-THEN rule links a condition, described using linguistic variables and fuzzy sets, to an output or action. The IF part captures the knowledge, while the THEN part provides the output. The fuzzy inference system widely uses these IF-THEN rules to compute the degree to which the input data matches the condition of a rule.

\noindent Fuzzy inference encompasses two methods: Mamdani and Sugeno, each of which is explained as follows:

\vspace{0.1cm}

\noindent \textbf{Mamdani Method:}

\noindent \noindent The Mamdani fuzzy inference system was proposed by Ebrahim Mamdani\cite{b13}. It was initially designed to control a steam engine and boiler combination using a set of linguistic control rules obtained from experienced human operators. In the Mamdani inference system, the output of each rule is a fuzzy set.

\vspace{0.1cm}

\noindent \textbf{Sugeno Method:}

\noindent This fuzzy inference system was proposed by Takagi, Sugeno, and Kang to develop a systematic approach for generating fuzzy rules from a given input-output dataset\cite{b14}.

\noindent A typical fuzzy rule in a first-order Sugeno fuzzy model has the form:

\hspace{1.1cm} IF $x$ is $A$ and $y$ is $B$ THEN $z = f(x, y)$

\vspace{0.1cm}

\noindent Where $A$ and $B$ are fuzzy sets in the antecedent, and 

\noindent $z = f(x, y)$ is a crisp function in the consequent.

\vspace{0.1cm}

\paragraph{Defuzzification}

\hspace{0.01cm} \newline
The defuzzification process is intended to convert the fuzzy output back into a crisp or classical output, aligning it with the control objective. Since the fuzzy conclusion or output remains a linguistic variable, it needs to be transformed into a crisp variable through the defuzzification process.

\noindent Three common defuzzification techniques found in literature are: the Mean of Maximum method, the Center of Gravity method, and the Height method \cite{b15}. 

\noindent In the context of Mamdani's fuzzy inference system, which generates fuzzy sets as output, defuzzification becomes necessary to convert the fuzzy output into a specific, crisp value.

\noindent In contrast, Sugeno inference systems typically provide a mathematical expression as output, thereby eliminating the explicit need for defuzzification, as the result is already in crisp form.

\subsection{Reinforcement Learning}

\noindent Reinforcement Learning (RL) is a subset of machine learning that empowers AI systems, often referred to as agents, to learn and adapt through a process of trial and error. In this paradigm, agents interact with their environment and receive feedback in the form of negative or positive signals, which are interpreted as punishments or rewards. The primary goal of an RL agent is to maximize the cumulative reward, often referred to as the reward function. \cite{b25}

\noindent This learning process allows RL agents to improve their performance over time by learning from their mistakes and successes. Through continuous interaction with the environment, the agent adjusts its actions to achieve better outcomes, effectively refining its strategy to maximize rewards. This iterative approach mimics aspects of natural intelligence, where learning and adaptation are driven by experience and feedback.

\noindent Reinforcement Learning is particularly powerful because it does not require explicit programming for every possible scenario. Instead, the agent autonomously discovers the optimal actions through exploration and exploitation. This capability makes RL a key component in developing advanced AI systems that can operate in dynamic and complex environments, such as robotics, game playing, and autonomous driving. By closely replicating the way natural intelligence learns and adapts, RL represents a significant advancement in the field of artificial intelligence.

\subsubsection{Value Functions and Bellman Equations}

\noindent When the agent acts given its state under the policy $\pi(a \mid s)$, the transition probability function $\mathcal{P}$ determines the subsequent state $s'$.
$$\mathcal{P}_{s s'}^a = \mathcal{P}(s' \mid s, a) = \mathbb{P}[S_{t+1} = s' \mid S_t = s, A_t = a]$$
When the agent acts based on its policy $\pi(a \mid s)$ and transitions to a new state determined by the transition probability function $\mathcal{P}_{s s'}^a$, it receives a reward based on the reward function as feedback:

$$\mathcal{R}_s^a = \mathbb{E}[\mathcal{R}_{t+1} \mid S_t = s, A_t = a]$$

\noindent Rewards are short-term, given as feedback after the agent takes an action and transitions to a new state. Summing all future rewards and discounting them leads to the return $\mathcal{G}$:
$$\mathcal{G}_t = \sum_{i=0}^N \gamma^i \mathcal{R}_{t+1+i}$$

\noindent Where $\gamma$ is the discount factor, ranging from 0 to 1.

\noindent With the return $\mathcal{G}$, we define the state-value function $\mathcal{V}_\pi$ (how good it is to be in a state) and the action-value or q-value function $\mathcal{Q}_\pi$ (how good it is to take an action, being in a state):
$$\mathcal{V}_\pi(s) = \mathbb{E}_\pi[\mathcal{G}_t \mid S_t = s]$$
$$\mathcal{Q}_\pi(s, a) = \mathbb{E}_\pi[\mathcal{G}_t \mid S_t = s, A_t = a]$$

\noindent Since our policy determines the agent's actions given its state, achieving an optimal policy $\pi_*$ means achieving optimal actions, which is our goal.

\noindent Optimal policy $\pi_*$ $\rightarrow$ optimal state-value and action-value functions $\rightarrow$ maximum return $\rightarrow$ argmax of value functions:
$$\pi_* = \arg\max_\pi \mathcal{V}_\pi(s) = \arg\max_\pi \mathcal{Q}_\pi(s, a)$$

\noindent To calculate the argmax of value functions, we need the maximum return $\mathcal{G}_t$, which requires the maximum sum of rewards $\mathcal{R}_s^a$. This is where the Bellman Equations come into play.

\noindent The Bellman Equations break down the value functions into two parts:
1. Immediate reward
2. Discounted future value function

\noindent The state-value function can be expressed as:
$$
\begin{aligned}
\mathcal{V}_\pi(s) & = \mathbb{E}[\mathcal{G}_t \mid S_t = s] \\
& = \mathbb{E}[\mathcal{R}_{t+1} + \gamma \mathcal{R}_{t+2} + \gamma^2 \mathcal{R}_{t+3} + \ldots \mid S_t = s] \\
& = \mathbb{E}[\mathcal{R}_{t+1} + \gamma (\mathcal{R}_{t+2} + \gamma \mathcal{R}_{t+3} + \ldots) \mid S_t = s] \\
& = \mathbb{E}[\mathcal{R}_{t+1} + \gamma \mathcal{G}_{t+1} \mid S_t = s] \\
& = \mathbb{E}[\mathcal{R}_{t+1} + \gamma \mathcal{V}_\pi(s_{t+1}) \mid S_t = s]
\end{aligned}
$$

\noindent The action-value function can also be expressed as:
$$\mathcal{Q}_\pi(s, a) = \mathbb{E}[\mathcal{R}_{t+1} + \gamma \mathcal{Q}_\pi(s_{t+1}, a_{t+1}) \mid S_t = s, A_t = a]$$

\subsubsection{Reinforcement Learning Algorithms}
RL encompasses algorithms aimed at enabling agents to learn optimal behaviors through interactions with their environment. These algorithms are categorized into several types: Value-Based Methods estimate value functions to derive optimal policies, with Q-Learning\cite{b16} and SARSA\cite{b17} exemplifying off-policy and on-policy approaches, respectively. Policy-Based Methods optimize policies directly without value functions, such as REINFORCE, suitable for continuous action spaces. Actor-Critic Methods integrate value and policy approaches, like A3C \cite{b18} , PPO \cite{b19} and DDPG \cite{b20}, to improve performance and stabilize training. Model-Based Methods use environment models for decision-making enhancement, Dyna-Q\cite{b22} combines model-free and model-based learning, and MBPO\cite{b23} optimizes policies with synthetic data for increased efficiency and performance.

\subsubsection{Twin Delayed Deep Deterministic Policy Gradient}

TD3 is an advanced actor-critic algorithm designed to address issues of overestimation bias and instability in reinforcement learning \cite{b21}. It extends the DDPG algorithm, which is used for continuous action spaces. TD3 introduces several key improvements:\vspace{0.1cm}

1. \textbf{Clipped Double Q-Learning:} TD3 uses two Q-networks (critics) to mitigate overestimation bias. The target value is the minimum of the two Q-values from the target networks, providing a more conservative estimate.
$$
y_t = r_t + \gamma \min_{i=1,2} Q_{\theta_i'}(s_{t+1}, \pi_{\phi'}(s_{t+1}) + \epsilon)
$$
\noindent Where \(\epsilon \sim \text{clip}(\mathcal{N}(0, \sigma), -c, c)\) is a small noise added for target smoothing.\vspace{0.1cm}

2. \textbf{Delayed Policy Updates:} The policy (actor) is updated less frequently than the Q-networks. Typically, the actor is updated every two iterations of the critic updates. This helps stabilize training and prevents the policy from being updated based on outdated Q-values.
$$
\theta_i \leftarrow \theta_i - \lambda_Q \nabla_{\theta_i} \frac{1}{N} \sum (Q_{\theta_i}(s_t, a_t) - y_t)^2
$$
\noindent Where \(N\) is the batch size, \(\theta_i\) are the parameters of the critic networks, and \(\lambda_Q\) is the learning rate for the critic.
$$
\phi \leftarrow \phi - \lambda_\pi \nabla_\phi \frac{1}{N} \sum Q_{\theta_1}(s, \pi_\phi(s))
$$
\noindent Where \(\phi\) are the parameters of the actor network, and \(\lambda_\pi\) is the learning rate for the actor.\vspace{0.1cm}

3. \textbf{Target Policy Smoothing:} To add robustness to the policy, TD3 applies noise to the target action, which is clipped to a certain range. This technique helps prevent the value function from exploiting errors in the policy by averaging over similar actions.
$$
\pi_{\phi'}(s_t) \leftarrow \pi_{\phi'}(s_t) + \epsilon
$$
\noindent Where \(\epsilon \sim \text{clip}(\mathcal{N}(0, \sigma), -c, c)\).

\section{Model and Control Strategy}

\subsection{Model Used for Simulation}
The management of Type 1 Diabetes Mellitus through automatic insulin injection systems necessitates robust mathematical models that simulate the intricate dynamics of the disease. 
Among the various mathematical models available, the basic Hovorka system has been widely adopted in artificial pancreas systems to facilitate the synthesis of control algorithms\cite{b8}. This model adeptly captures the input-output relationships inherent in diabetes management, where the primary inputs include subcutaneous insulin infusion and, periodically, meal ingestion and intravenous glucose infusion. The outputs are primarily the intravenous glucose concentrations.

\subsubsection{Glucose Subsystem}
The glucose subsystem is modeled with a two-compartment approach\cite{b7}:
\begin{align}
\frac{dQ_1(t)}{dt} = &-\left(\frac{F_{c01}}{V_G G(t)} + x_1(t)\right) Q_1(t) + k_{12} Q_2(t) \notag \\
                     &- F_R + U_G(t) + EGP_0[1 - x_3(t)]  \\
\frac{dQ_2(t)}{dt} = &x_1(t) Q_1(t) - (k_{12} + x_2(t)) Q_2(t)
\end{align}

Where \(Q_1\) and \(Q_2\) are the glucose masses in the accessible and non-accessible compartments, respectively, and \(k_{12}\) is the transfer rate constant. \(V_G\) is the distribution volume of the accessible compartment, \(G\) is the glucose concentration, and \(EGP_0\) represents the endogenous glucose production.

\paragraph{Non-Insulin-Dependent Glucose Flux and Renal Clearance}
\begin{align}
F_{c01} &= \begin{cases} 
F_{01}, & \text{if } G \geq 4.5 \text{ mmol L}^{-1} \\
\frac{F_{01} G}{4.5}, & \text{otherwise}
\end{cases}, \\
F_R &= \begin{cases}
0.003 (G - 9) V_G, & \text{if } G > 9 \text{ mmol L}^{-1} \\
0, & \text{otherwise}
\end{cases}.
\end{align}

\paragraph{Gut Absorption Rate}
\begin{equation}
U_G(t) = \frac{D_G A_G t e^{-t/t_{max,G}}}{t_{max,G}^2},
\end{equation}
where \(t_{max,G}\) is the time-of-maximum glucose appearance, \(D_G\) is the digested carbohydrates amount, and \(A_G\) is the bioavailability.

\subsubsection{Insulin Subsystem}
\paragraph{Insulin Absorption}
Insulin absorption is modeled as\cite{b7}:
\begin{align}
\frac{dS_1(t)}{dt} &= u(t) - \frac{S_1(t)}{t_{max,I}}, \\
\frac{dS_2(t)}{dt} &= \frac{S_1(t)}{t_{max,I}} - \frac{S_2(t)}{t_{max,I}},
\end{align}
where \(S_1\) and \(S_2\) represent insulin in two compartments of subcutaneously administered insulin.

\paragraph{Plasma Insulin Concentration}
\begin{equation}
\frac{dI(t)}{dt} = \frac{U_I(t)}{V_I} - k_e I(t),
\end{equation}
where \(k_e\) is the fractional elimination rate and \(V_I\) is the distribution volume.

\subsubsection{Insulin Action Subsystem}
The insulin action on glucose kinetics is modeled through\cite{b7}:
\begin{align}
\frac{dx_1}{dt} &= -k_{a1} x_1(t) + k_{b1} I(t), \\
\frac{dx_2}{dt} &= -k_{a2} x_2(t) + k_{b2} I(t), \\
\frac{dx_3}{dt} &= -k_{a3} x_3(t) + k_{b3} I(t),
\end{align}
where \(x_1\), \(x_2\), and \(x_3\) represent insulin's effects on glucose distribution/transport, disposal, and endogenous glucose production, respectively.

\subsection{The fuzzy logic controller}
For our fuzzy controller, we have adopted a Type 1 Takagi-Sugeno fuzzy inference
system, a choice driven by its suitability for optimization tuning. The Takagi-Sugeno
model, while sacrificing interpretability, aligns well with our optimization-focused approach.
Our fuzzy inference system features two inputs: insulin level and insulin rate or speed,
each with three membership functions (trapezoidal, triangular, and trapezoidal) The tuning of these membership functions was conducted manually through simulations, guided
by intuition and experience. This configuration results in a total of 9 rules, yielding 27
parameters for tuning.
\[ u_i = a_i e + b_i \dot{e} + c_i, \quad \text{for } i = 1 \text{ to } 9, \]

 where $u_i$ is the fuzzified controller output.\\
The output employs a linear membership function, and for defuzzification, we employ the weighted average method.

\subsection{RL Environment}
The Reinforcement Learning environment in the context of artificial pancreas systems involves creating a simulated setting that closely mimics the physiological processes of glucose regulation in humans. This environment serves as the training ground for the RL agent to learn optimal control strategies for insulin administration. 
\subsubsection{Observations}
The observations that were fed to the RL agent from the environment included current blood glucose levels and their rate of change. It's worth noting that these observations undergo a normalization process. Normalization is employed to prevent the emergence of excessively large or excessively small parameters, thereby enhancing numerical stability within the system. 

\subsubsection{The reward function}
The reward function is designed based on the error $e(t)$  we carefully considered key criteria derived from the glucose levels of a type 1 diabetes patient. The normal glucose range, defined as 80 to 100 mg/dL, serves as a crucial reference point. Within this range, the RL agent receives a significantly higher reward to reinforce behaviors that maintain glucose levels within the healthy boundaries. We also designated an acceptable range, consisting of two sub-ranges (100 to 160 mg/dL and 70 to 80 mg/dL), where the agent receives a moderate reward. This encourages the agent to navigate glucose levels within an acceptable but broader spectrum. Additionally, we identified two distinct bad ranges (160 or higher and 0 to 70 mg/dL) and assigned a negative reward within these intervals.we also considered the derivative $de(t)$ to guide the reinforcement learning agent toward desired performance. Additionally, $i(t)$ the integral of the absolute value of the normalized error, aimed at mitigating large overshooting and oscillations. Furthermore, $c(t)$ denotes the integral of the control signal, serving to penalize the agent for excessive insulin usage. The reward is given to the agent at each step. And the formulation is as follows:

\begin{itemize}
    \item If the error is less than 90 (the glucose is less than the reference), the reward will be given by the linear function: $(1-\lvert e \rvert )/20$

    \item If the error is greater than 90 (the glucose is greater than the reference), the reward will be given by the linear function: $(1-\lvert e \rvert )/70$

    \item If the error is less than 10 (the glucose is close to our reference), the reward will be given by the function, which at \(e=0\) gives the value 20, and at \(e=10\) gives 0:  $1.262 \cdot \left(\lvert e \rvert\right)^{\frac{1}{5}} + 2$

\item We then apply penalties to our agent as follows: $-2 \times 10^{-6} \cdot i - 10^{-6} \cdot c$, where the first term is introduced to mitigate oscillations in the blood glucose signal, ensuring stability, and the second term is designed to optimize the injected insulin, discouraging excessive usage. These penalties contribute to fine-tuning the agent's behavior for more effective and stable blood glucose regulation.
\end{itemize}
\subsubsection{Stop of episode}
In our specific implementation, episodes conclude under three conditions: after reaching the predefined simulation time of 24 hours, or when the glucose level exceeds 300 (indicating hyperglycemia), or falls below 50 (indicating hypoglycemia). These extreme glucose levels are considered hazardous situations for the patient, demanding immediate intervention. In such instances, we assert that the agent has failed to maintain glucose within the desired range, prompting a termination of the episode. This safety-centric approach aligns with our commitment to avoiding potentially harmful glucose extremes and ensures that the reinforcement learning agent operates within clinically acceptable bounds during the artificial pancreas control simulations.

\section{Experimentation and Results}
We present a thorough and comprehensive comparative analysis of three distinct and well-defined methods for controlling blood glucose levels, specifically tailored to address the challenges associated with Type 1 diabetes management. The methods evaluated in this study include a direct reinforcement learning 
agent, which autonomously learns optimal control strategies; a non-adaptive fuzzy logic controller with finely tuned and optimized parameters; and an adaptive fuzzy logic controller, where the parameters are continuously and dynamically adjusted by the RL agent at each simulation timestep.

\vspace{0.1cm}

\noindent It is important to explicitly highlight that all simulations, as well as the tuning processes for each of the proposed methods, were systematically carried out using MATLAB2022b and Simulink.

\subsection{Direct insulin control through Reinforcement Learning }

\noindent The results of the Direct insulin control through Reinforcement Learning in the nominal
case are presented below:
\begin{figure}[htbp]
    \centering
    \includegraphics[width=0.9\linewidth]{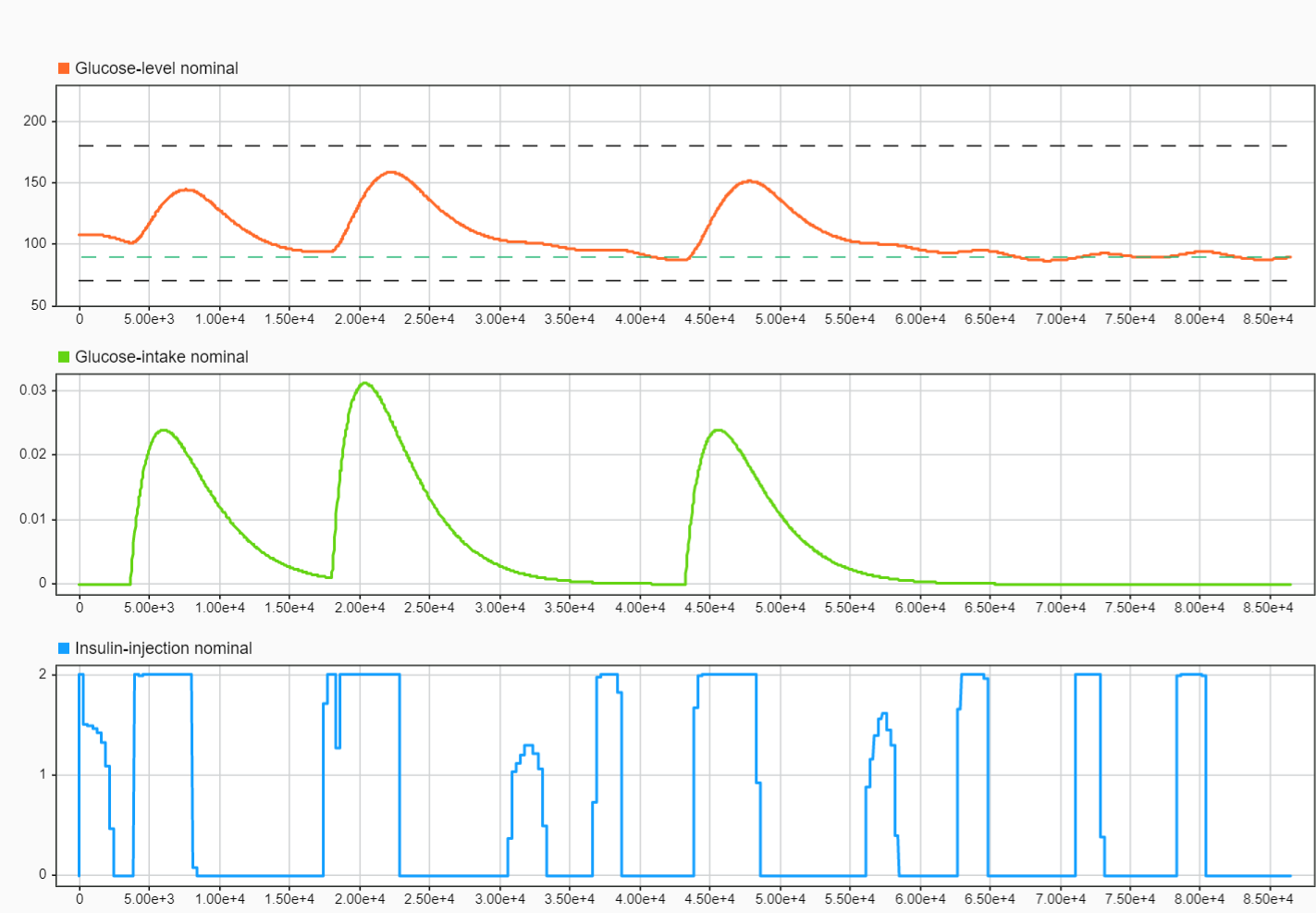}
    \caption{Glucose level (mg/dL) response of the Direct insulin control through Reinforcement Learning, along with corresponding meals intake (mg/dl) and insulin injections in units in the nominal case.}
    \label{fig:c2}
\end{figure}

\noindent In the Glucose-level subplot, the two black dotted lines clearly indicate the boundaries of the safe range for blood glucose levels, spanning from 70 to 180 units, while the green dotted line represents our chosen reference value, set at 90 units. It is important to emphasize that this visual representation, including both the safe range and the reference line, will consistently appear in all the glucose-level subplots presented throughout this article to maintain clarity and uniformity.

\vspace{0.2cm}

\noindent In order to thoroughly examine the robustness of the proposed approach, the Reinforcement Learning Controller was subjected to a comprehensive series of tests conducted under various sets of randomly generated time spans and differing quantities of meal intakes, allowing us to assess its performance across a wide range of possible scenarios.

\begin{figure}[htbp]
\centering
    \includegraphics[width=0.4\textwidth]{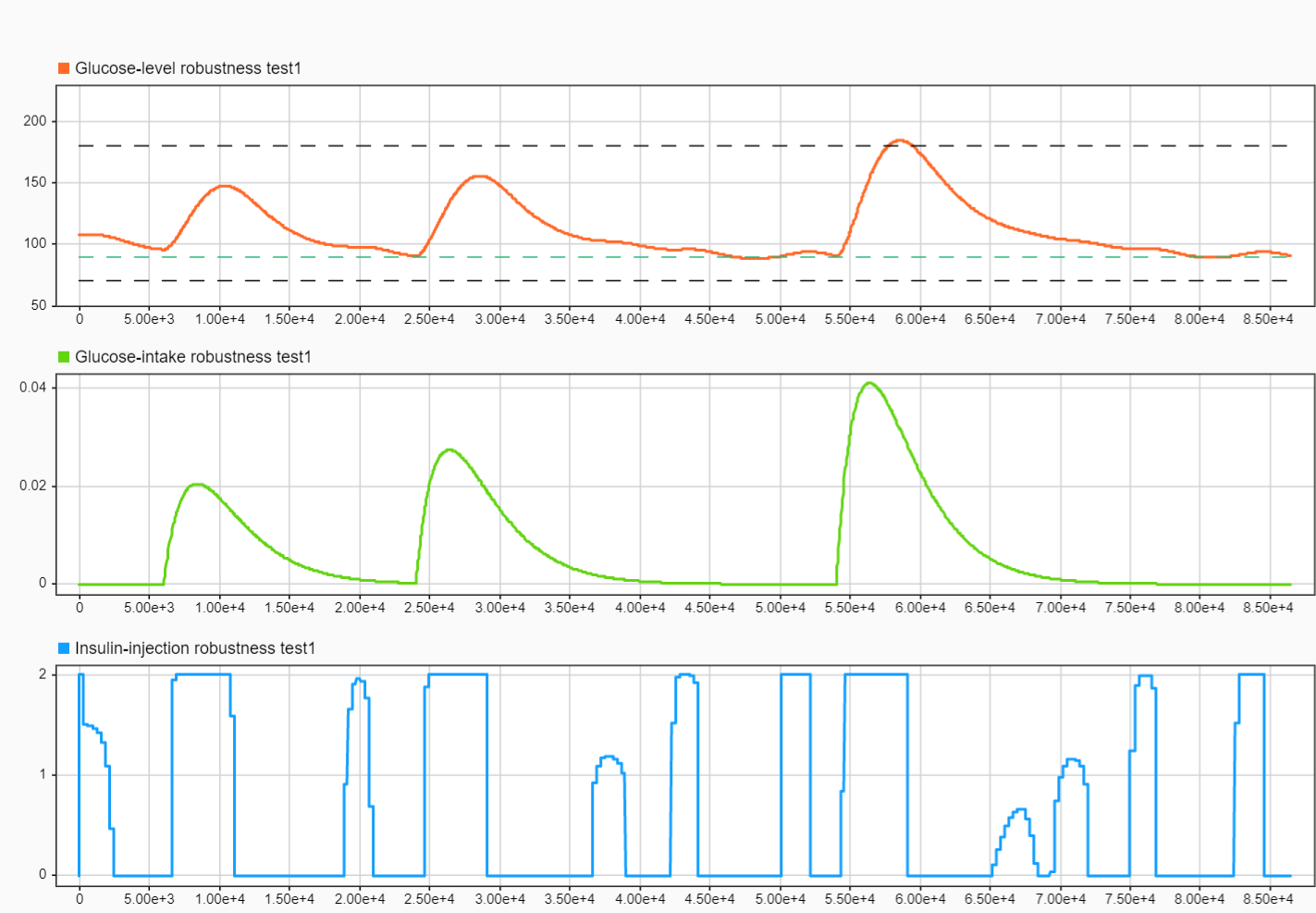}\\
    Case 1
    
    \vspace{1em} 
    
    \includegraphics[width=0.4\textwidth]{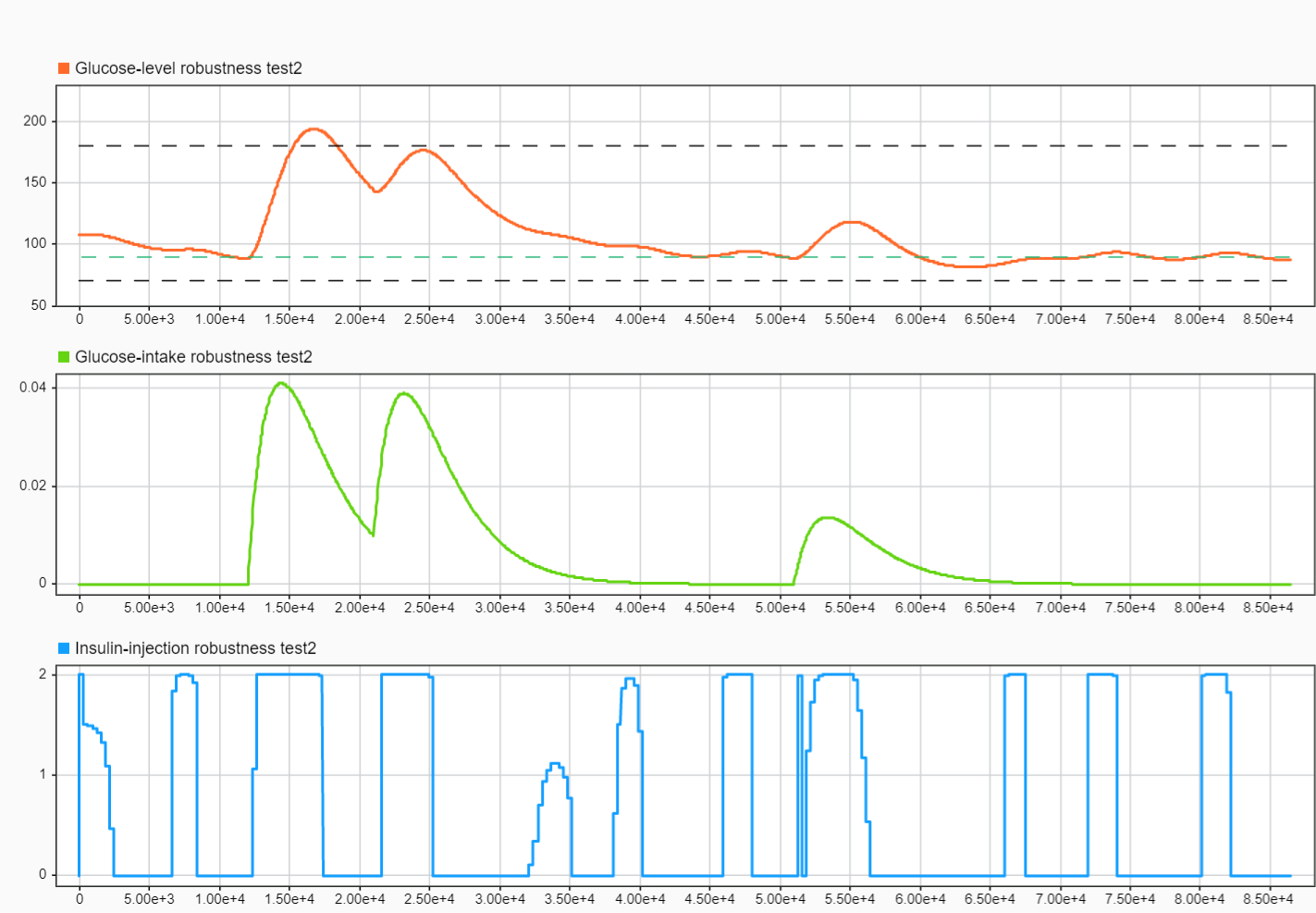}\\
     Case 2

    \vspace{1em} 

    \includegraphics[width=0.4\textwidth]{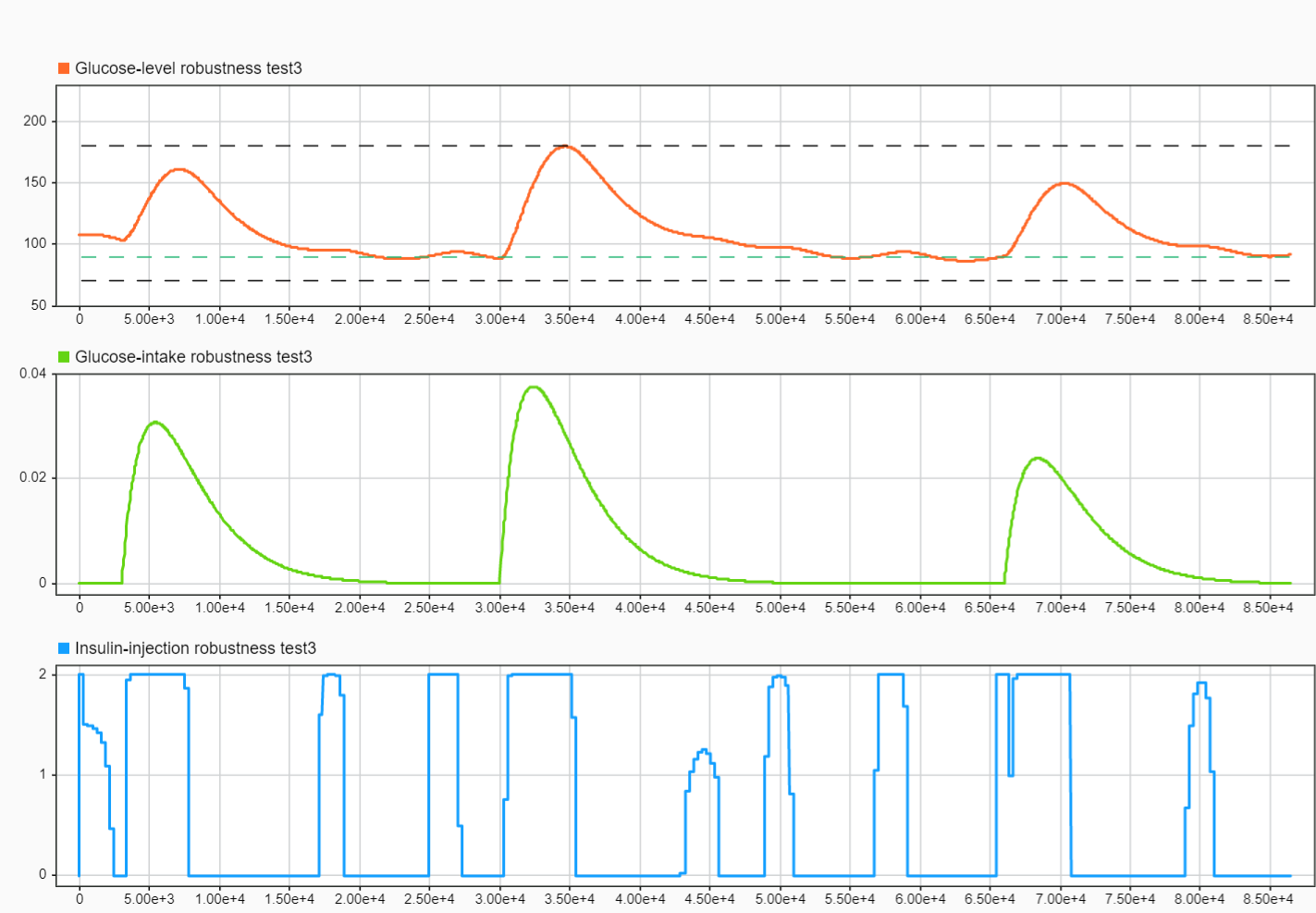}\\
    Case 3

    \vspace{1em} 

    \includegraphics[width=0.4\textwidth]{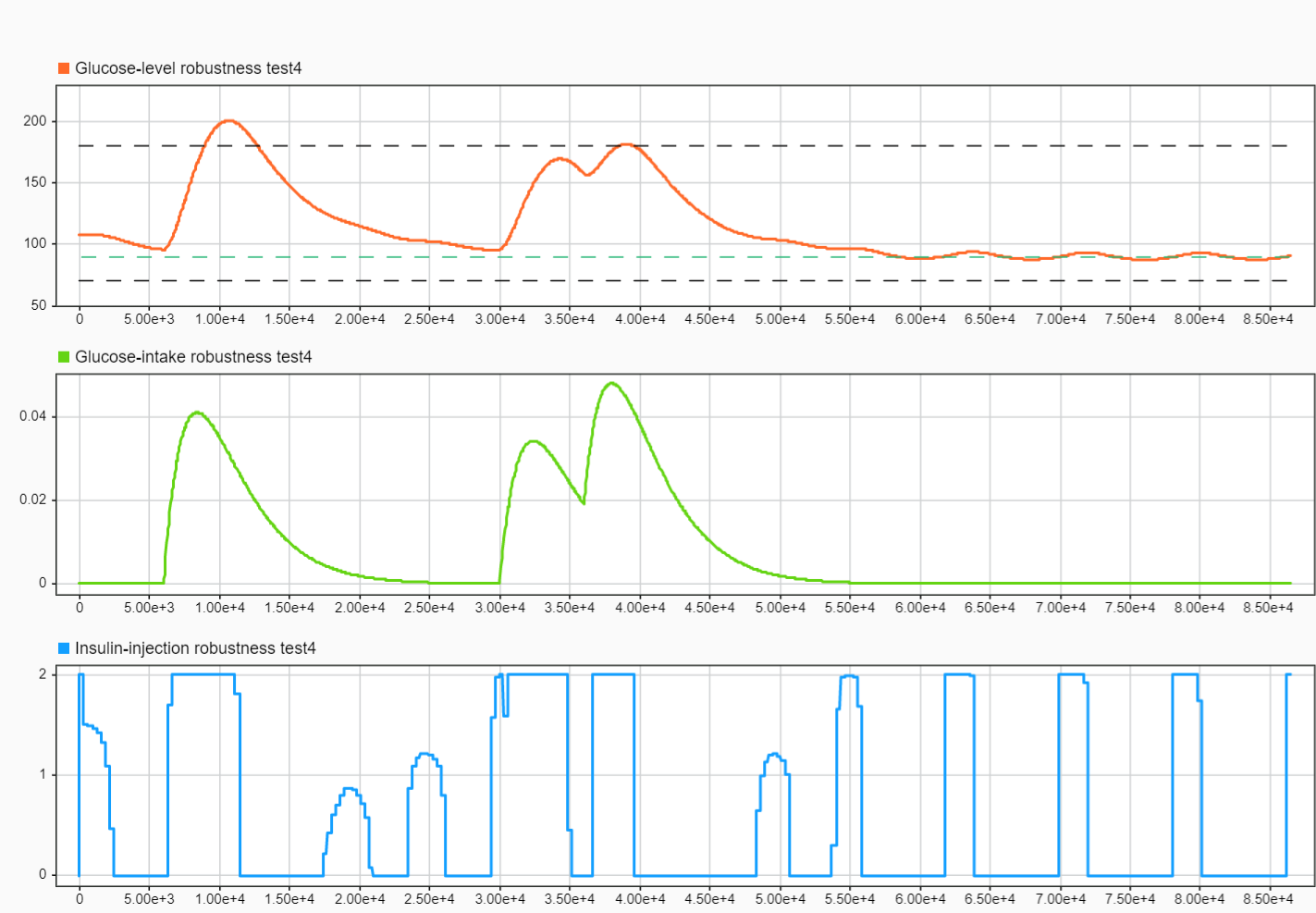}\\
    Case 4

\caption{Glucose level (mg/dL) response of the Direct insulin control through Reinforcement
Learning, along with corresponding meals intake (mg/dL) and insulin injections in units in the nominal
case.}
\label{fig: mat_fig}    
\end{figure}

\newpage

\subsection{Non-adaptive optimized fuzzy control}
The results obtained from the Non-Adaptive Takagi-Sugeno FLC under the nominal case scenario are presented below:\
 
\begin{figure}[htbp]
    \centering
    \includegraphics[width=0.9\linewidth]{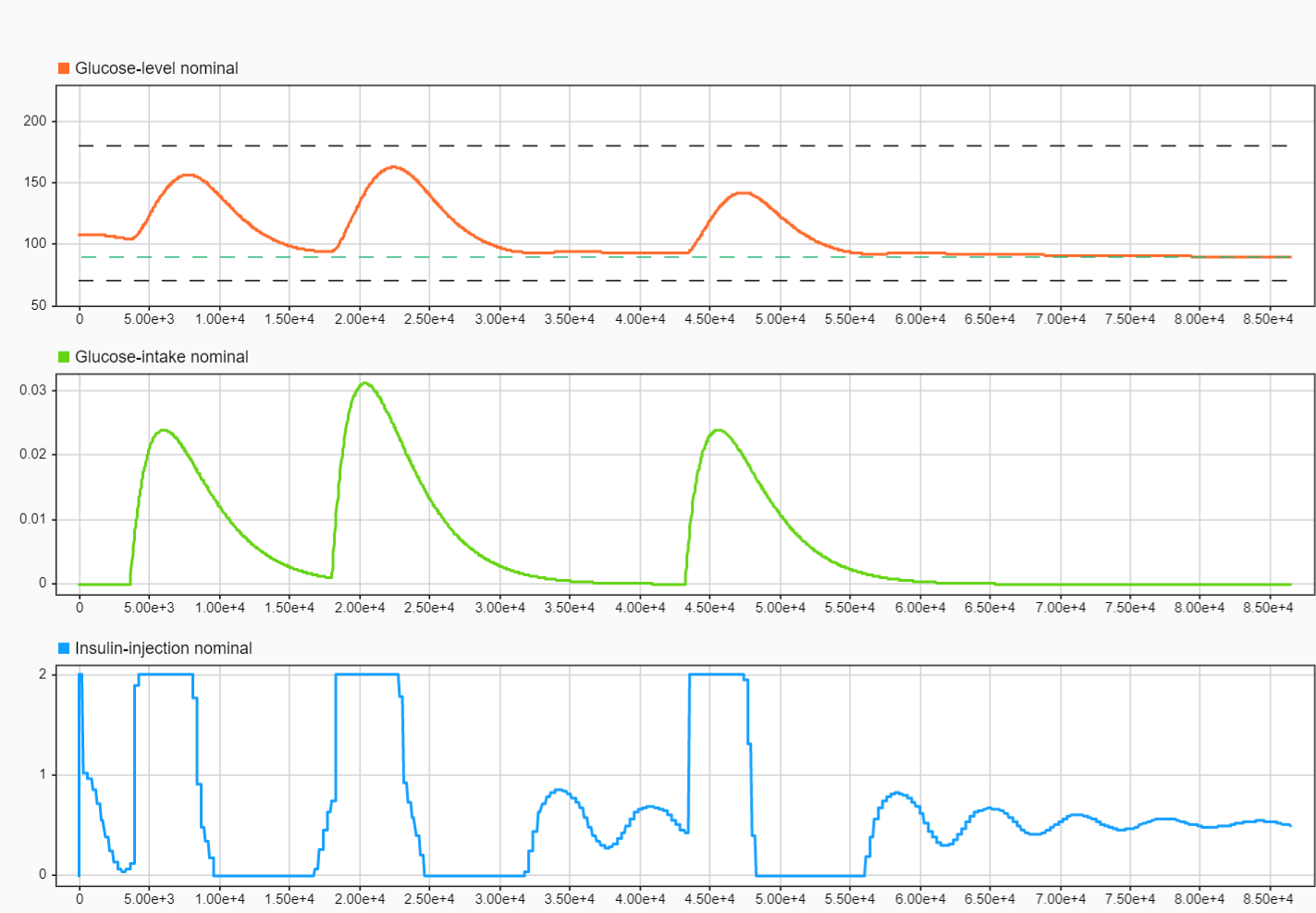}
    \caption{Glucose level (mg/dL) response of the Non-adaptive optimized fuzzy control, along with corresponding meals intake (mg/dL) and insulin injections in units in the nominal case.}
    \label{fig:enter-label}
\end{figure}

To thoroughly assess the robustness, the Non-Adaptive Takagi-Sugeno Fuzzy Logic Controller was tested across a variety of different sets of meals, each varying in timing and quantity.\\

\begin{figure}[htbp]
    \centering
    \includegraphics[width=0.4\textwidth]{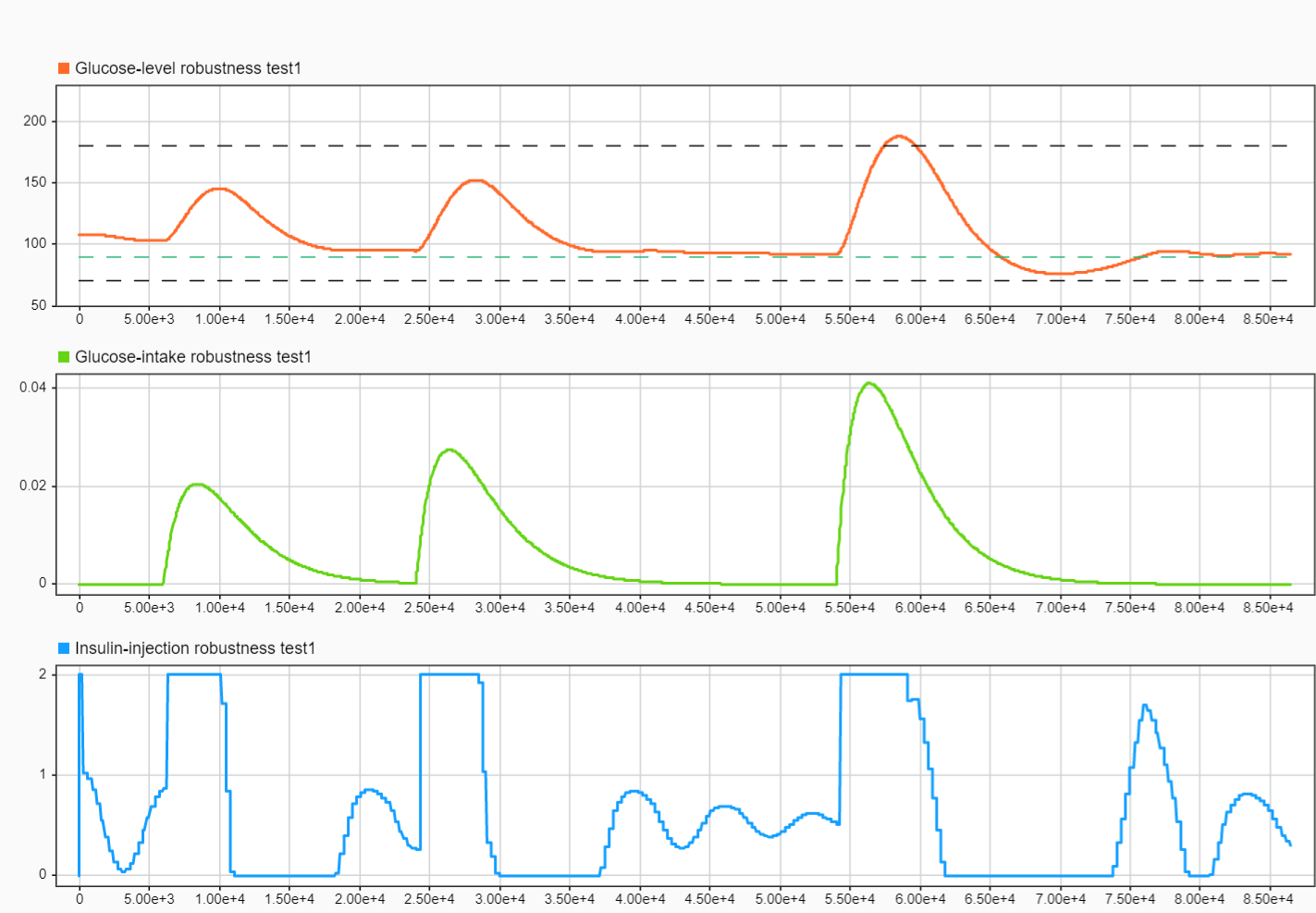}\\
    Case 1
\end{figure}

\begin{figure}[htbp]
    \centering
    \includegraphics[width=0.4\textwidth]{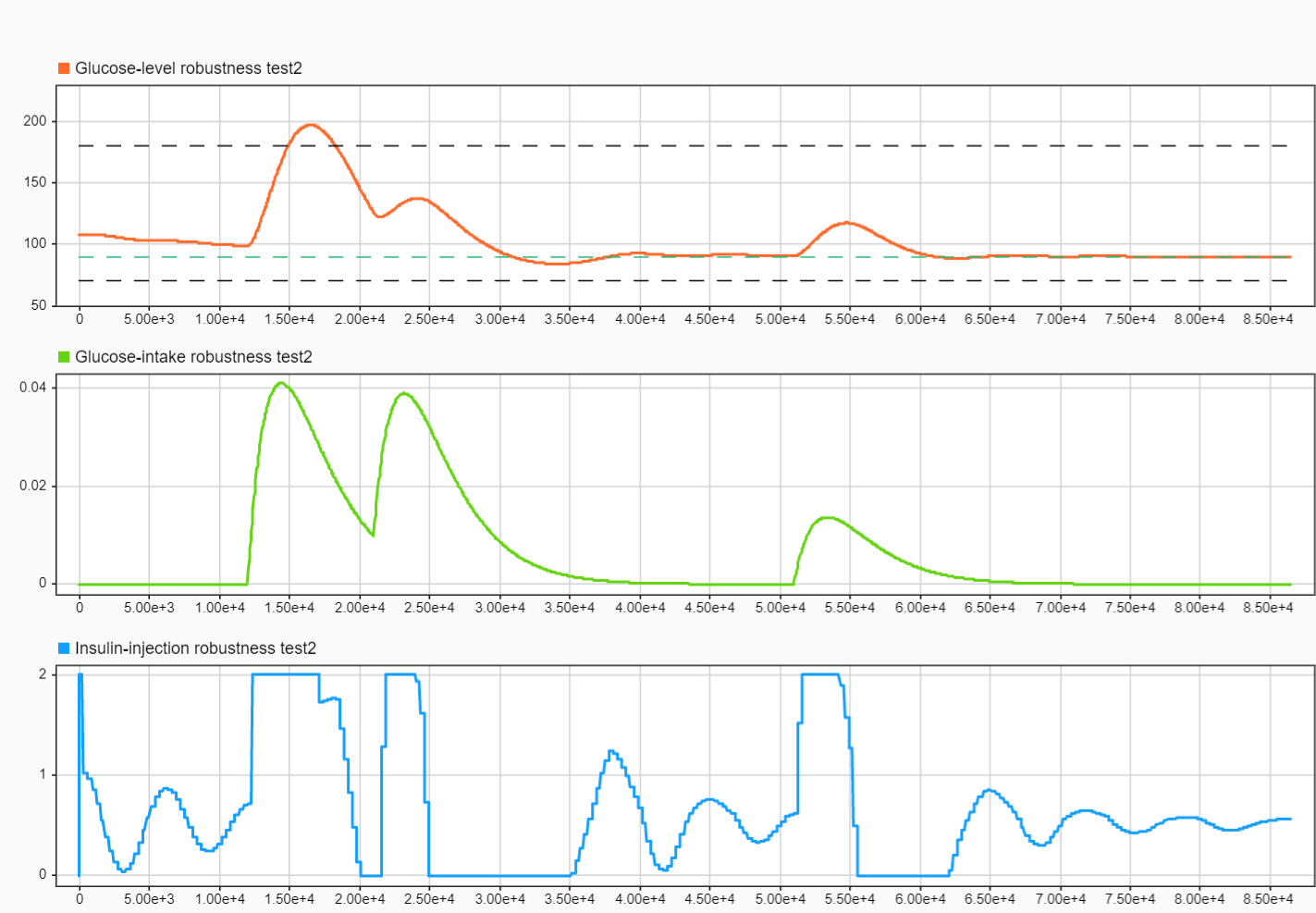}\\
    Case 2
\end{figure}

\begin{figure}[htbp]
    \centering
    \includegraphics[width=0.4\textwidth]{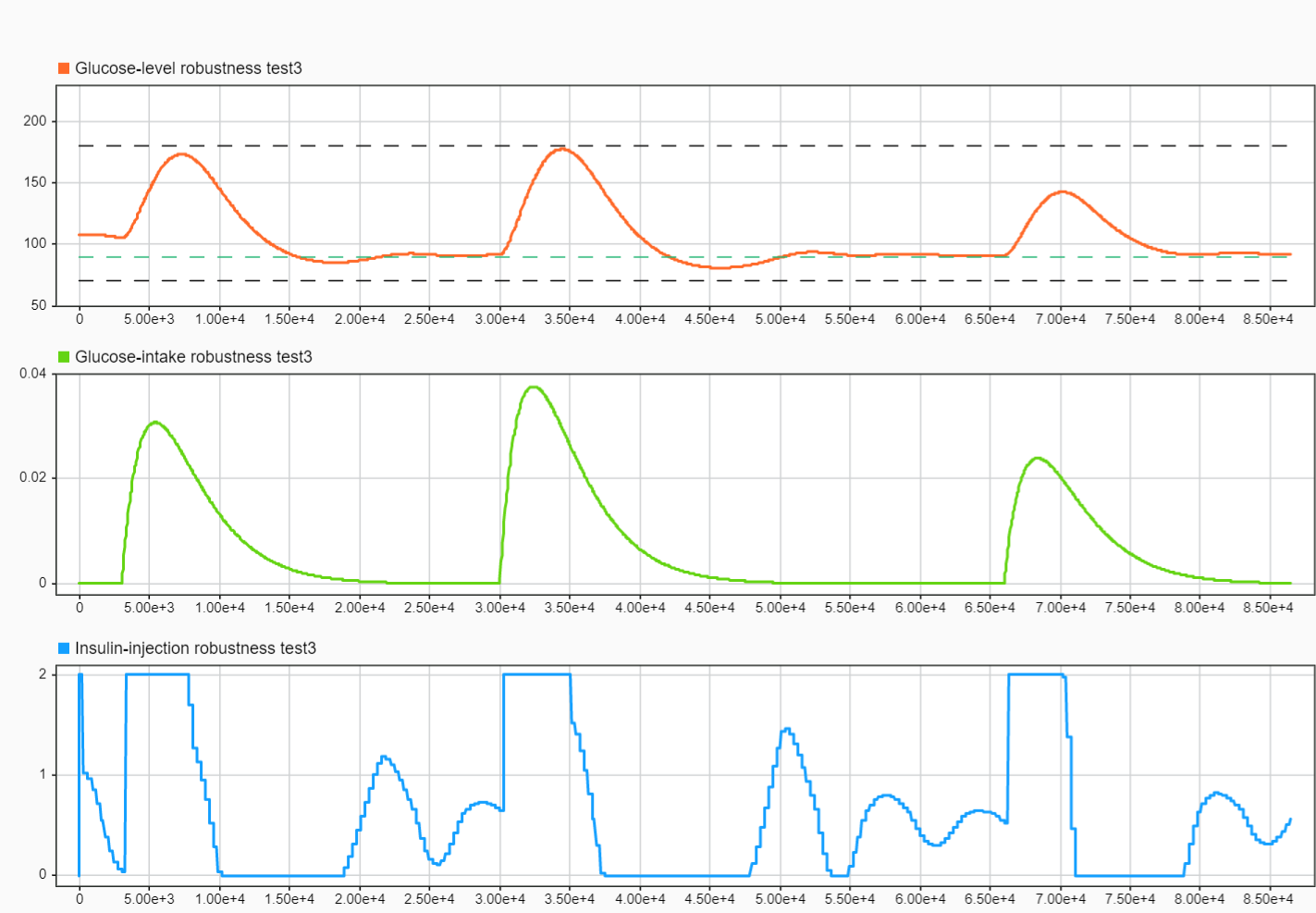}\\
    Case 3
\end{figure}

\begin{figure}[htbp]
    \centering
    \includegraphics[width=0.4\textwidth]{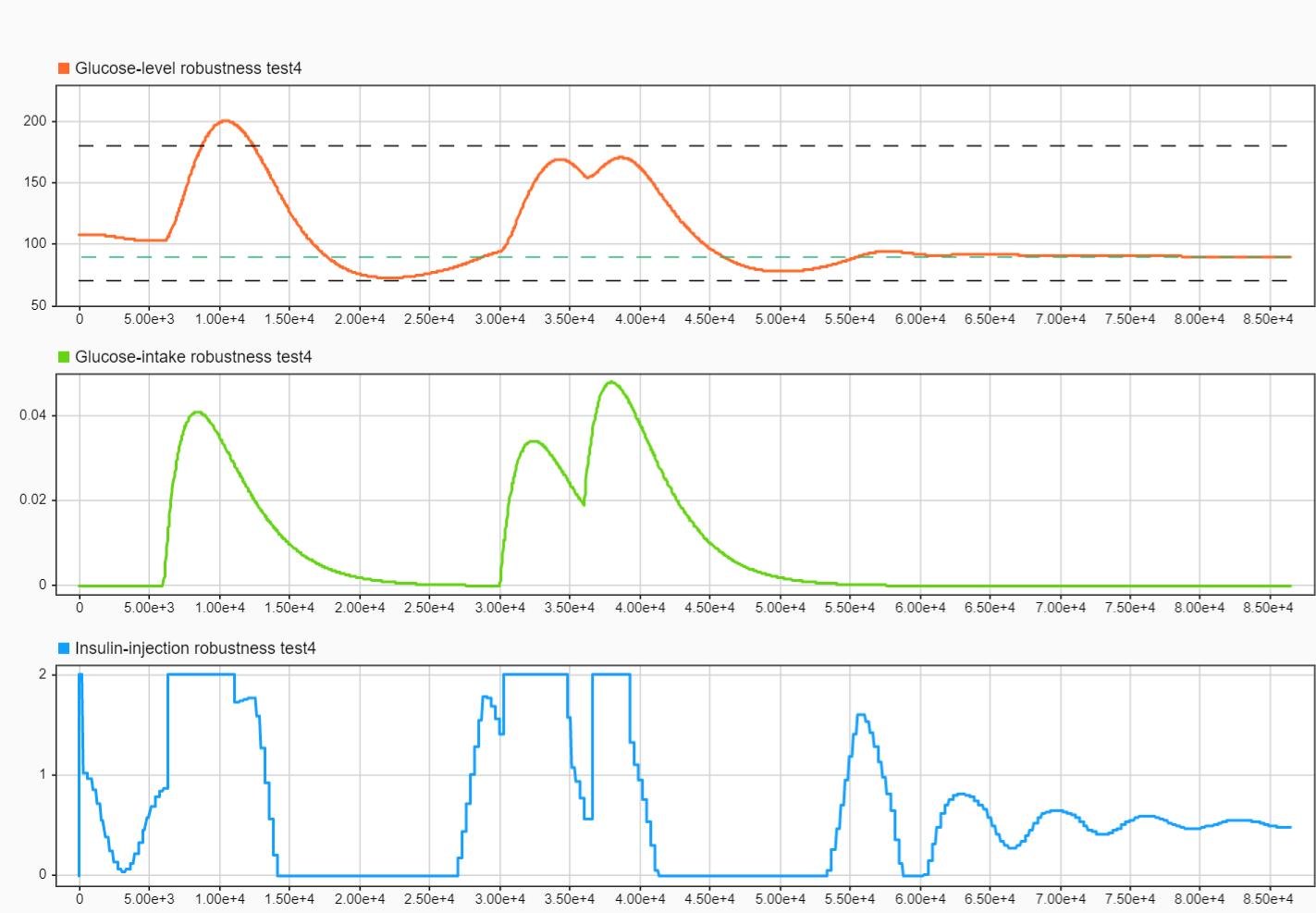}\\
    Case 4
    \caption{Glucose level (mg/dL) response of the Non-adaptive optimized fuzzy control, along with corresponding meals intake (mg/dL) and insulin injections in units in the nominal case.}
    \label{fig: mat_fig}
\end{figure}

\vspace{0.2cm}

\newpage
\subsection{Adaptive optimized fuzzy control}
 The results of the Adaptive Takagi-Sugeno FLC in the nominal case are presented below :\\
 
 To study the robustness, the Adaptive TS Fuzzy Logic Controller was tested under various sets meals
\begin{figure}[htbp]
    \centering
    \includegraphics[width=0.9\linewidth]{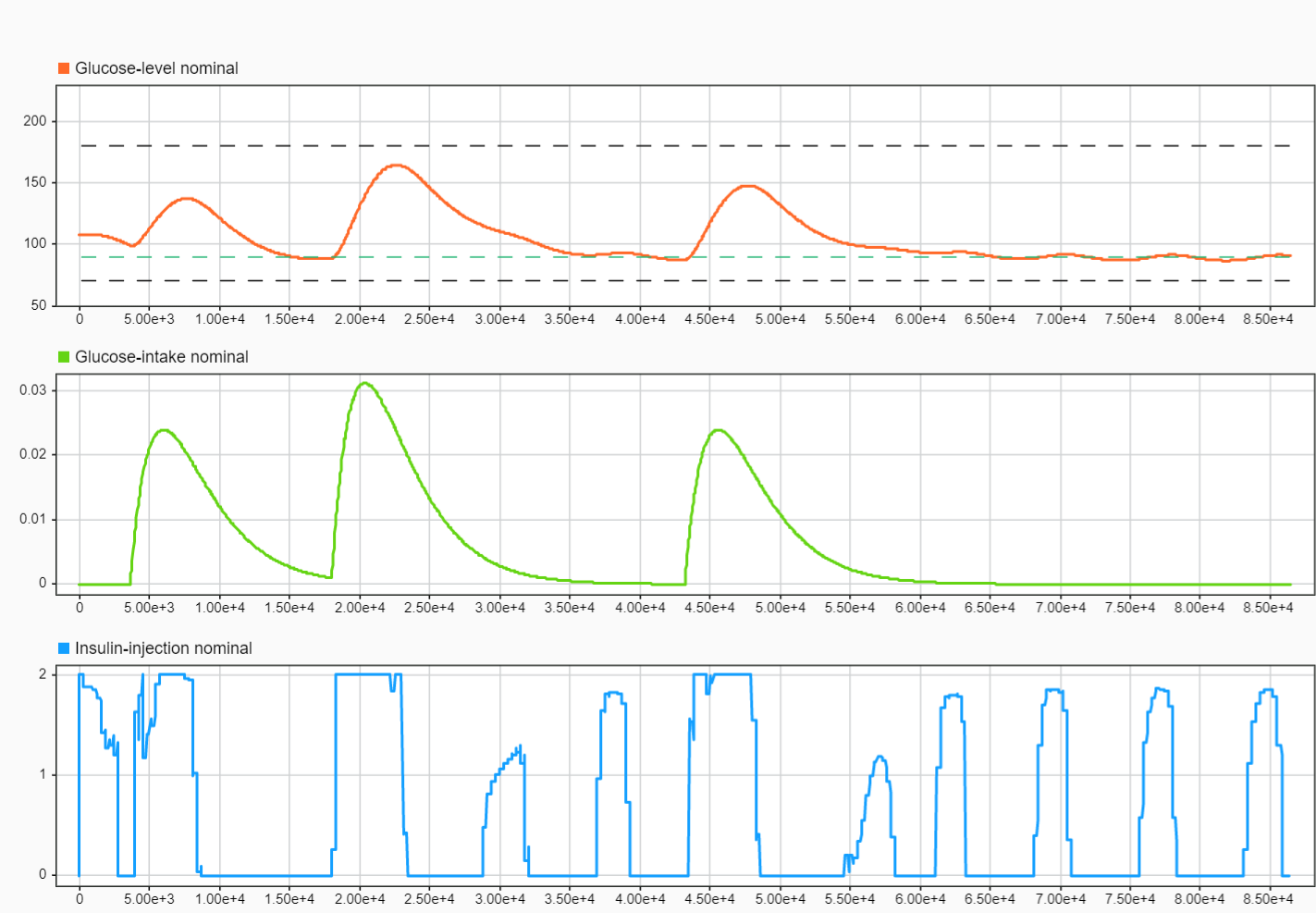}
    \caption{Glucose level (mg/dL) response of the adaptive FLC along with corresponding meals intake (mg/dL) and insulin injections in units in the nominal case.}
    \label{fig:enter-label}
\end{figure}

\begin{figure}[htbp]
\centering
    \includegraphics[width=0.4\textwidth]{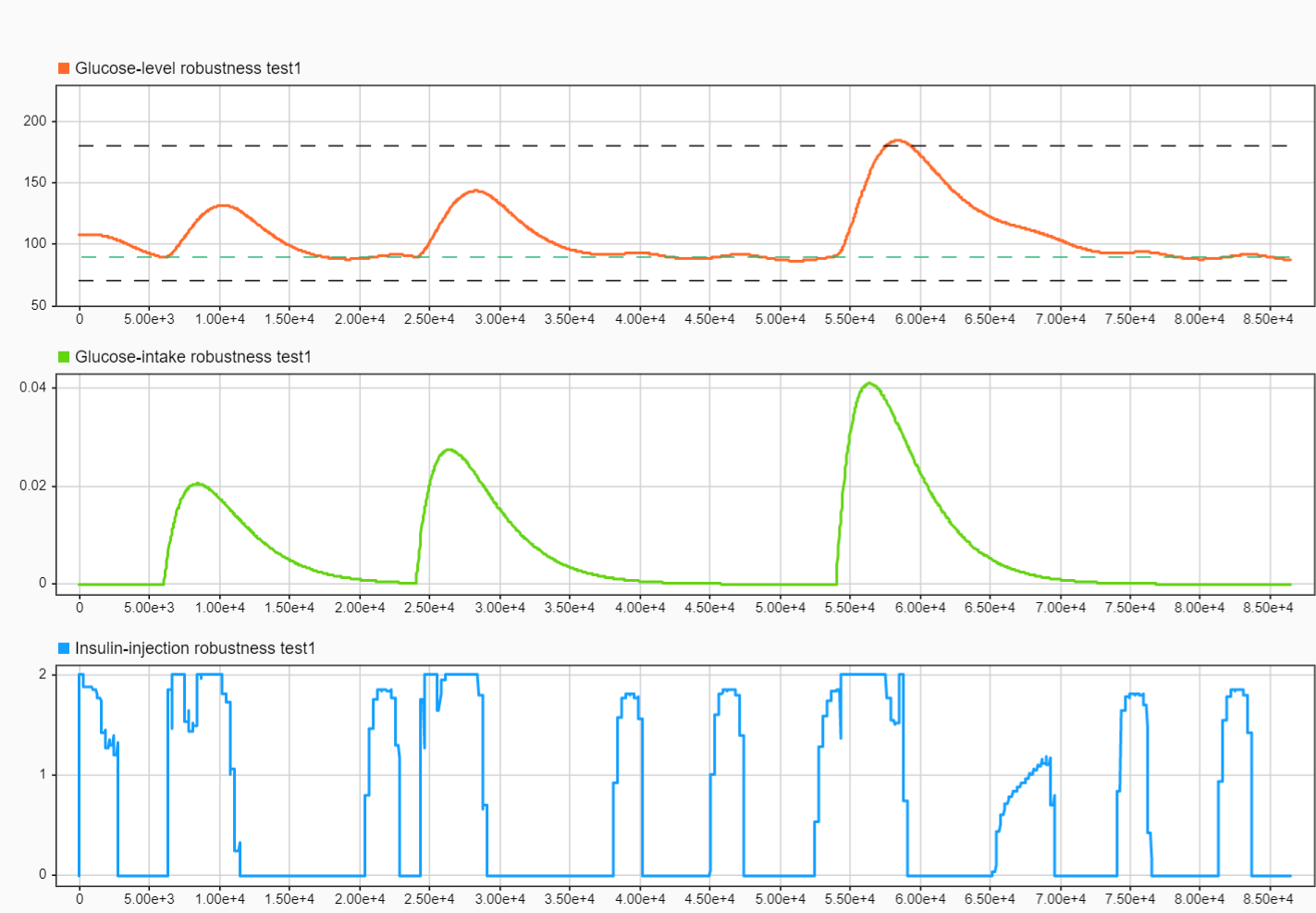}\\
    Case 1
    
    \vspace{1em} 
    
    \includegraphics[width=0.4\textwidth]{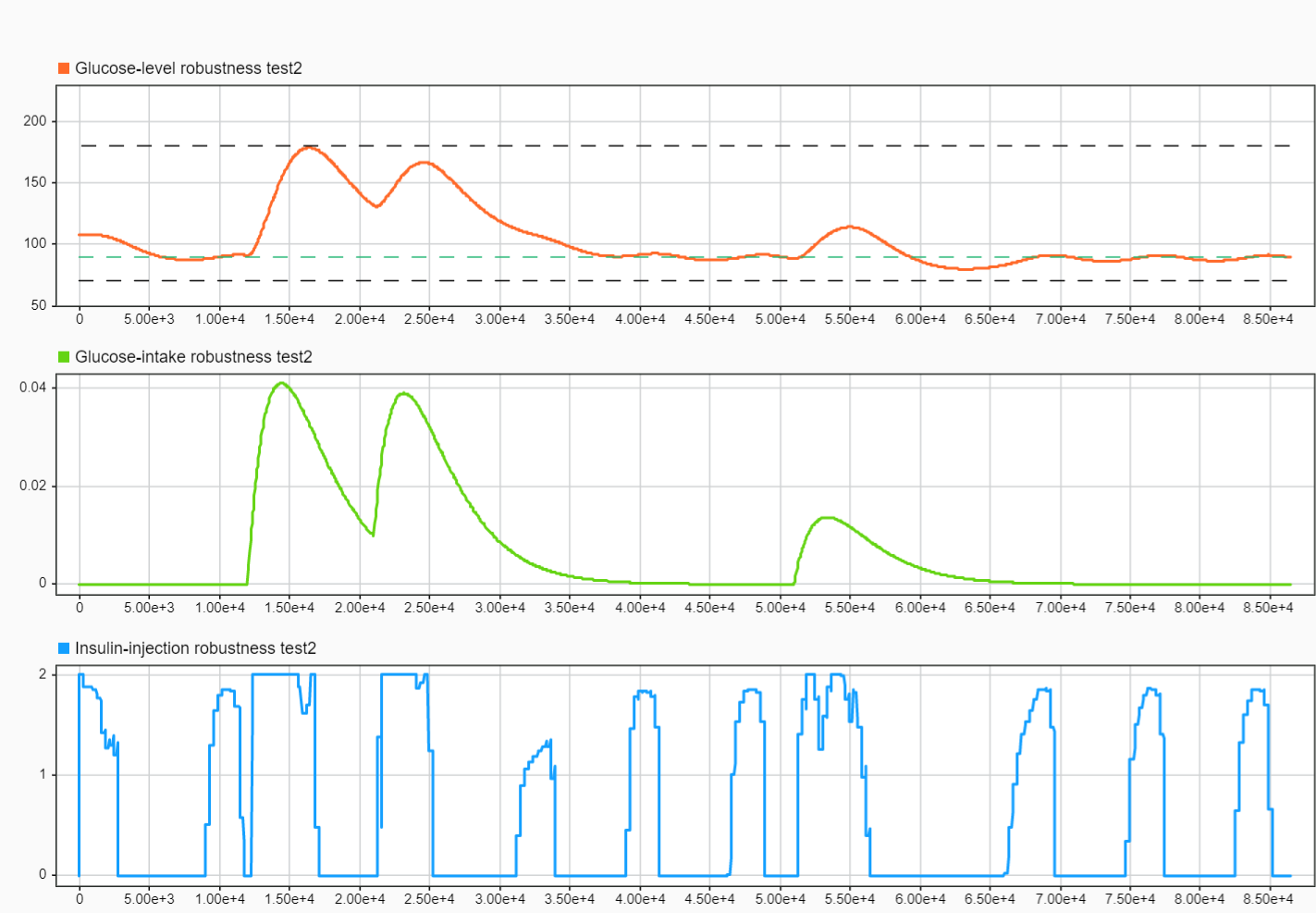}\\
    Case 2

    \vspace{1em} 

    \includegraphics[width=0.4\textwidth]{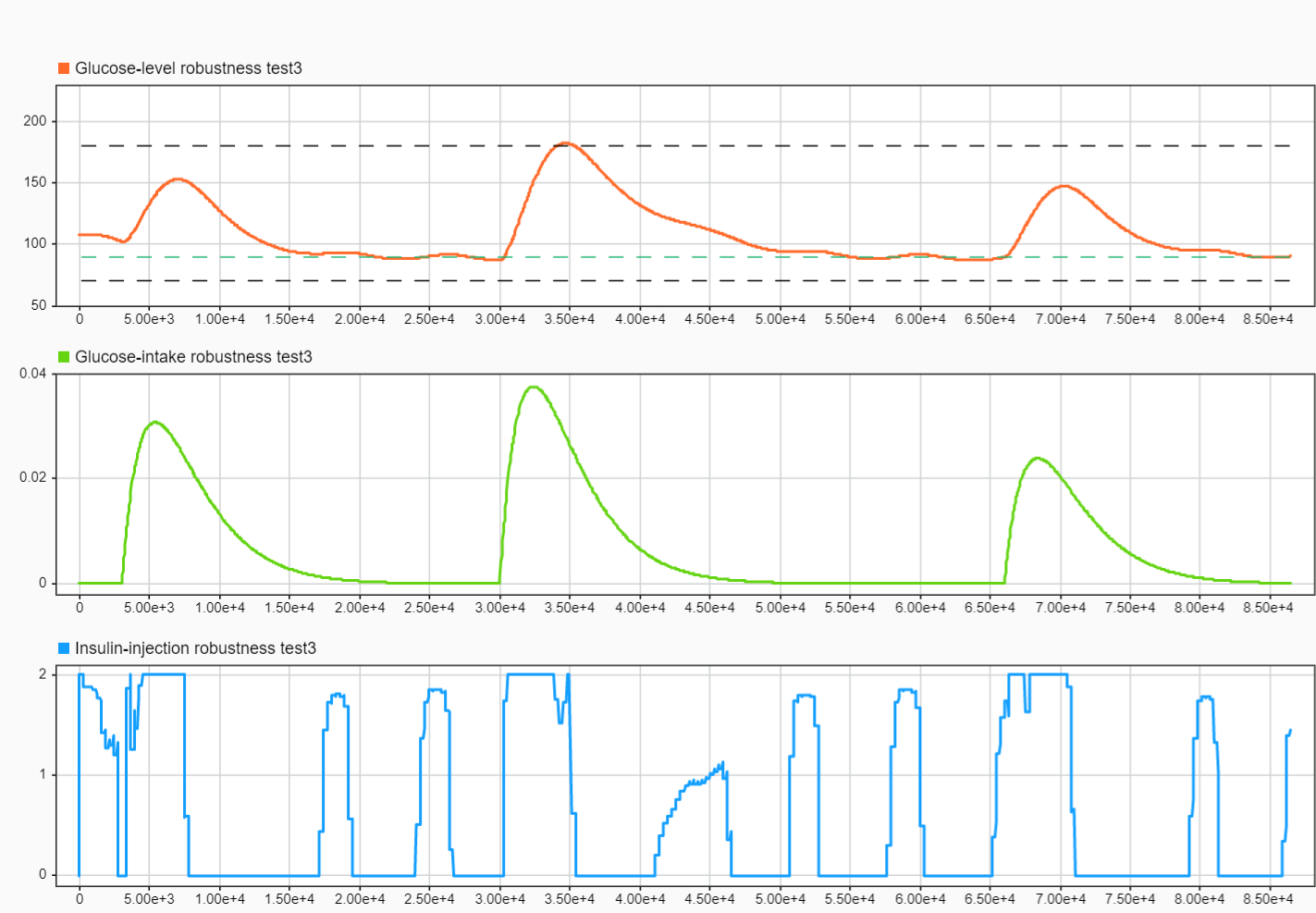}\\
    Case 3

    \vspace{1em} 

    \includegraphics[width=0.4\textwidth]{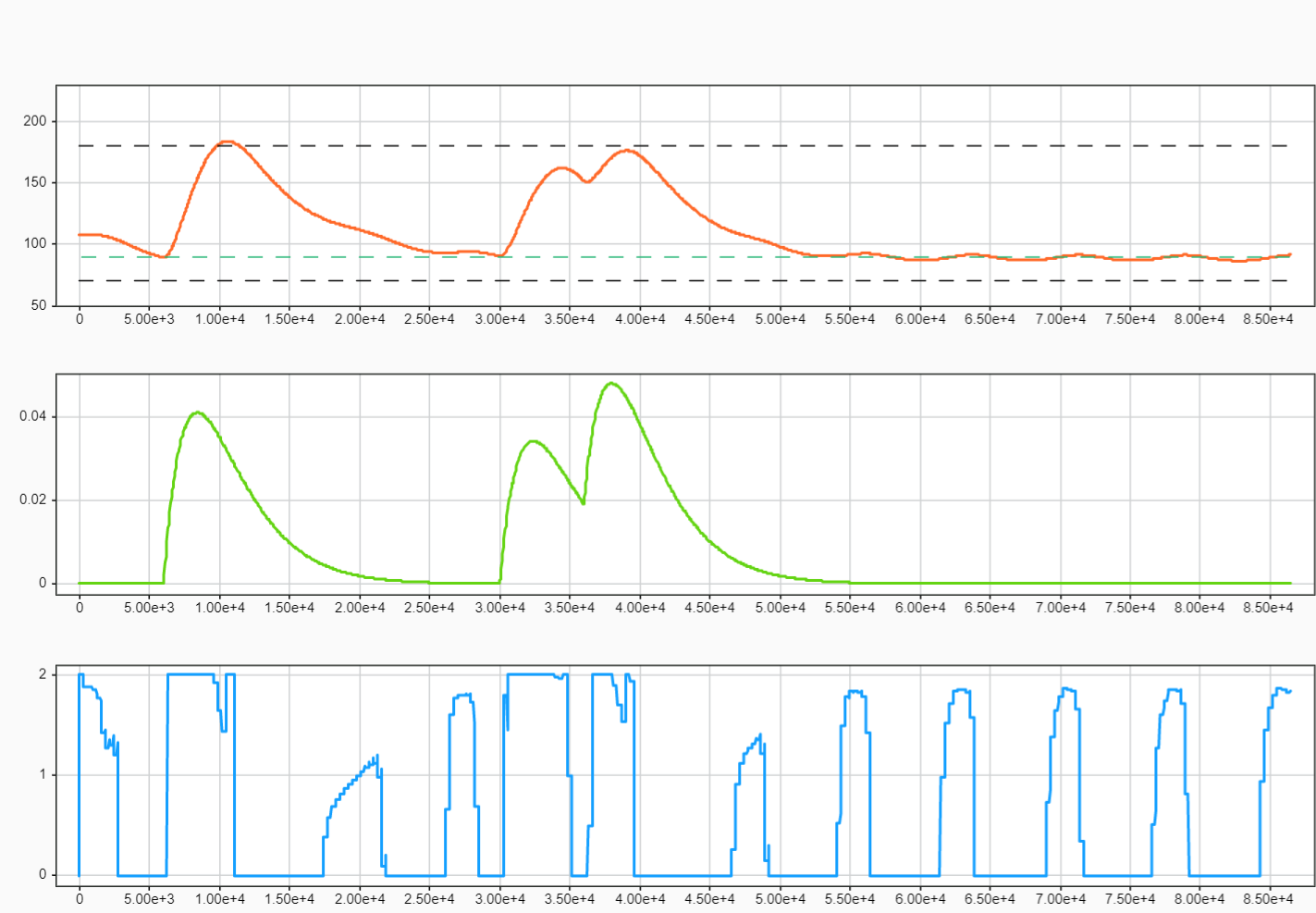}\\
    Case 4

\caption{Glucose level (mg/dL) response of the adaptive FLC, along with corresponding meals intake (mg/dL) and insulin injections in units in the nominal case.}
\label{fig: mat_fig}    
\end{figure}

\vspace{0.2cm}

\newpage
\subsection{Discussions}
The experimental results of our study, comparing three different controllers in both nominal and non-nominal cases, have provided valuable insights into their performances under various scenarios.

The first controller, based on direct insulin control through a reinforcement learning agent, demonstrated impressive results in the nominal case. Its robustness to different scenarios of meal times and quantities showcased good perturbation rejection capabilities. The adaptability of the RL agent allowed it to effectively handle variations in the system, making it a promising choice for dynamic environments. Unfortunately, the fact that the reinforcement learning agent is a black box makes it challenging to study its stability unlike classical approaches, which complicates placing trust in such applications.

On the other hand, the second controller, employing non-adaptive fuzzy controller, also performed well in the nominal case. However, its performance faltered when tested under different scenarios. Despite this, the non-adaptive fuzzy control approach is interesting. Given that we can determine the parameters of the fuzzy controller, it opens avenues for studying stability and robustness, ensuring safety and guarantee in unknown situations. This characteristic is particularly crucial in medical applications, such as artificial pancreas systems, where reliability is paramount.

To delve deeper into the controllers' behavior under challenging conditions, we propose a study examining the glucose response of both the direct reinforcement learning controller and the non-adaptive fuzzy logic controller in extreme cases. This will provide valuable insights into how each controller reacts and adapts to these scenarios.

The following figures shows the response of the Non-Adaptive Fuzzy Logic Controller and the Direct Reinforcement Learning Controller, both tested under extreme cases.

\begin{figure}[htbp]
    \centering
    \includegraphics[width=0.9\linewidth]{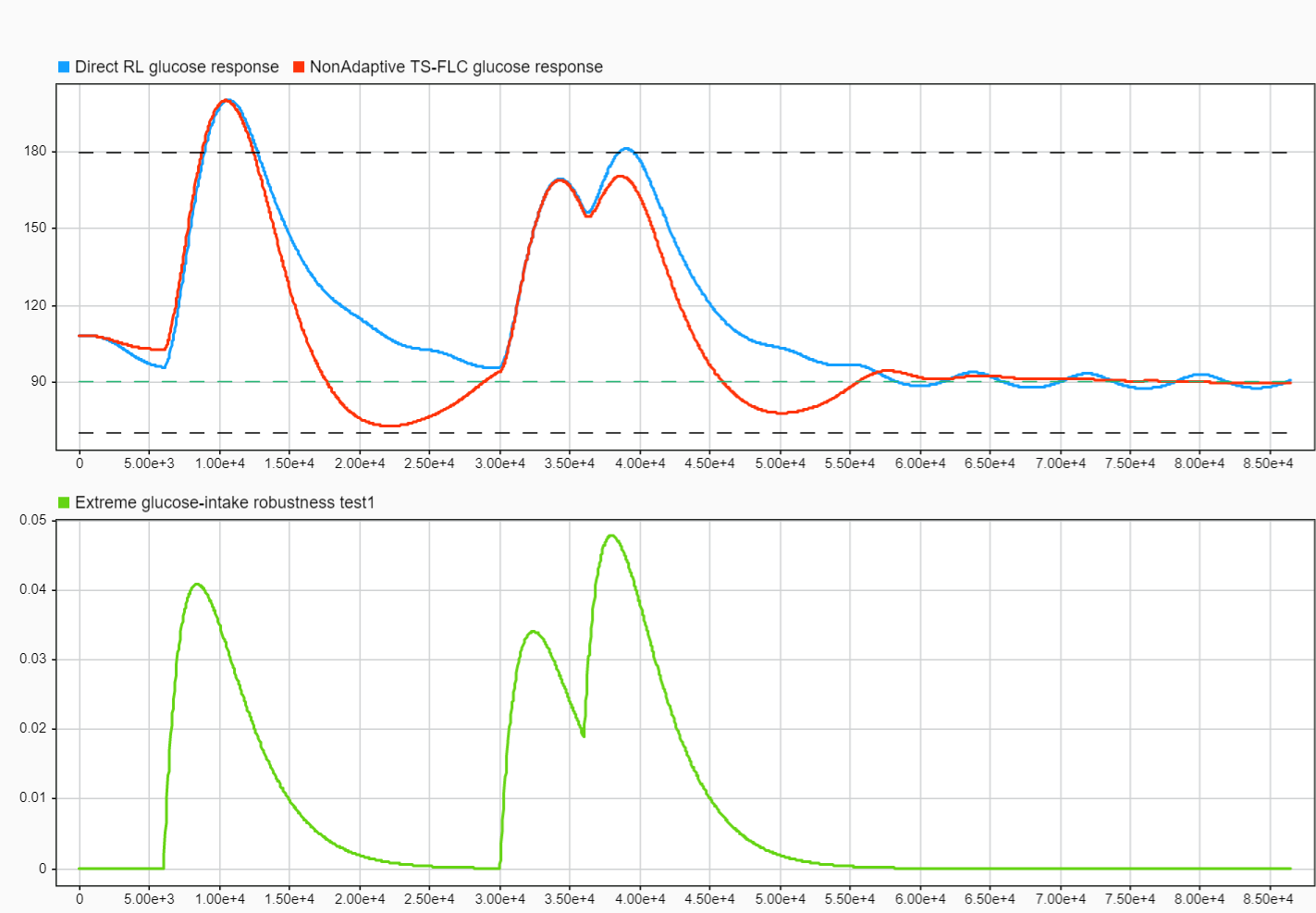}
    \caption{Glucose level (mg/dL) response of the Direct insulin control through Reinforcement Learning and the NonAdaptive Fuzzy Logic Controller, along with corresponding meals intake (mg/dL).}
    \label{fig:c2}
\end{figure}
We can see from these tests that when a very large glucose intake is taken, the NonAdaptive fuzzy logic controller, in its attempt to counter the glucose dynamics, undershoots the target value of 90. In contrast, the direct reinforcement learning controller does not undershoot the value 90. \\
\begin{figure}[htbp]
    \centering
    \includegraphics[width=0.9\linewidth]{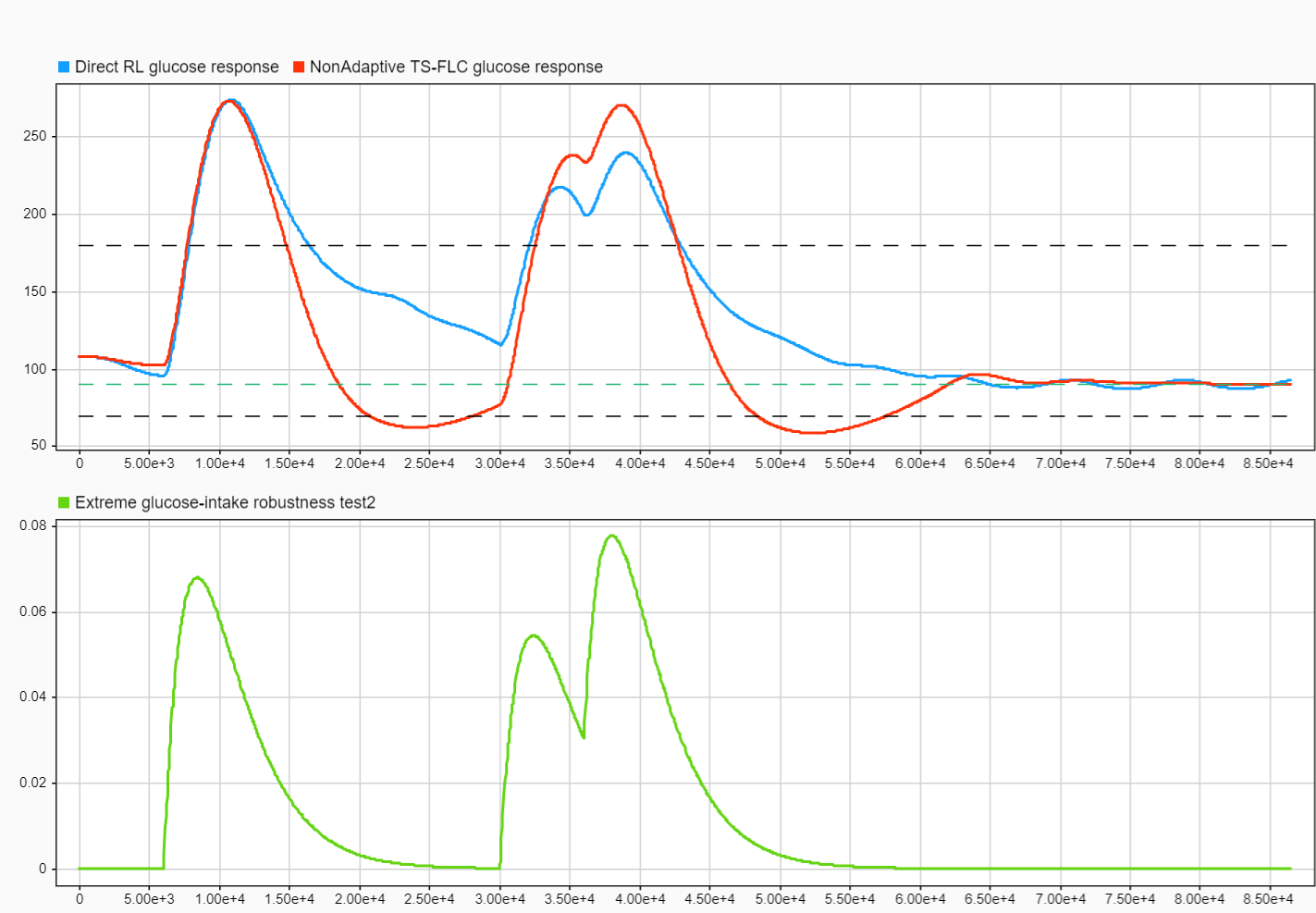}
    \caption{Glucose level (mg/dL) response of the Direct insulin control through Reinforcement Learning and the NonAdaptive Fuzzy Logic Controller, along with corresponding meals intake (mg/dL).}
    \label{fig:c2}
\end{figure}

To address the strengths of both approaches, we proposed an  adaptive fuzzy controller. This controller incorporates the adaptability of an RL agent while leveraging the stability and safety features of a fuzzy controller. The adaptive fuzzy controller dynamically tunes its parameters at each time step using the RL agent. This combination resulted in the most performant controller, excelling in both the nominal case and robustness tests.

The following figures shows the response of the Direct Reinforcement Learning Controller, the Non-Adaptive Fuzzy Logic Controller and the Adaptive Fuzzy Logic Controller, All tested under extreme cases.\\

\begin{figure}[htbp]
    \centering
    \includegraphics[width=0.9\linewidth]{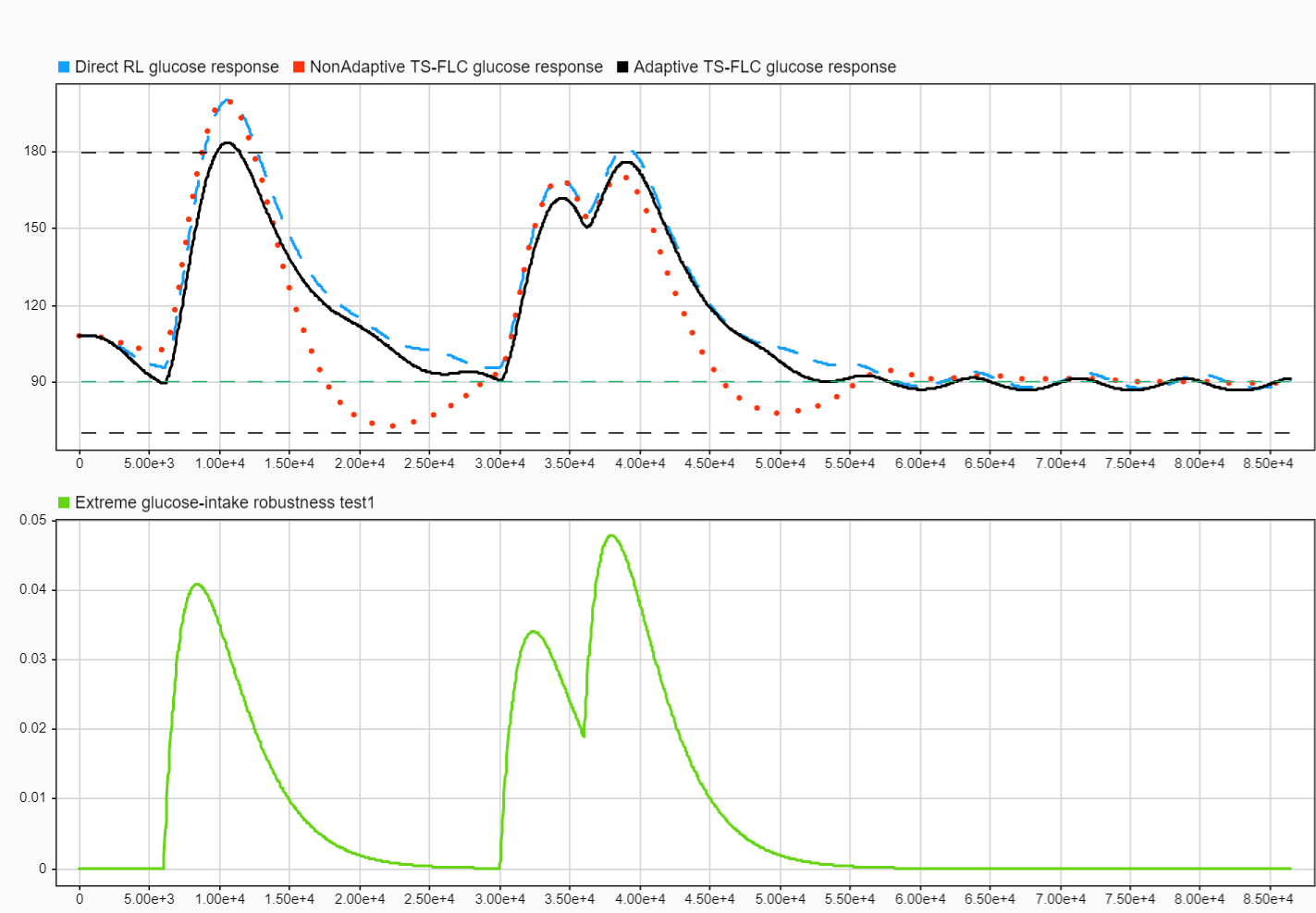}
    \caption{Glucose level (mg/dL) response of the Direct insulin control through Reinforcement Learning, the NonAdaptive Fuzzy Logic Controller and the Adaptive Fuzzy Logic Controller, along with corresponding meals intake(mg/dL).}
    \label{fig:c2}
\end{figure}

\begin{figure}[htbp]
    \centering
    \includegraphics[width=0.9\linewidth]{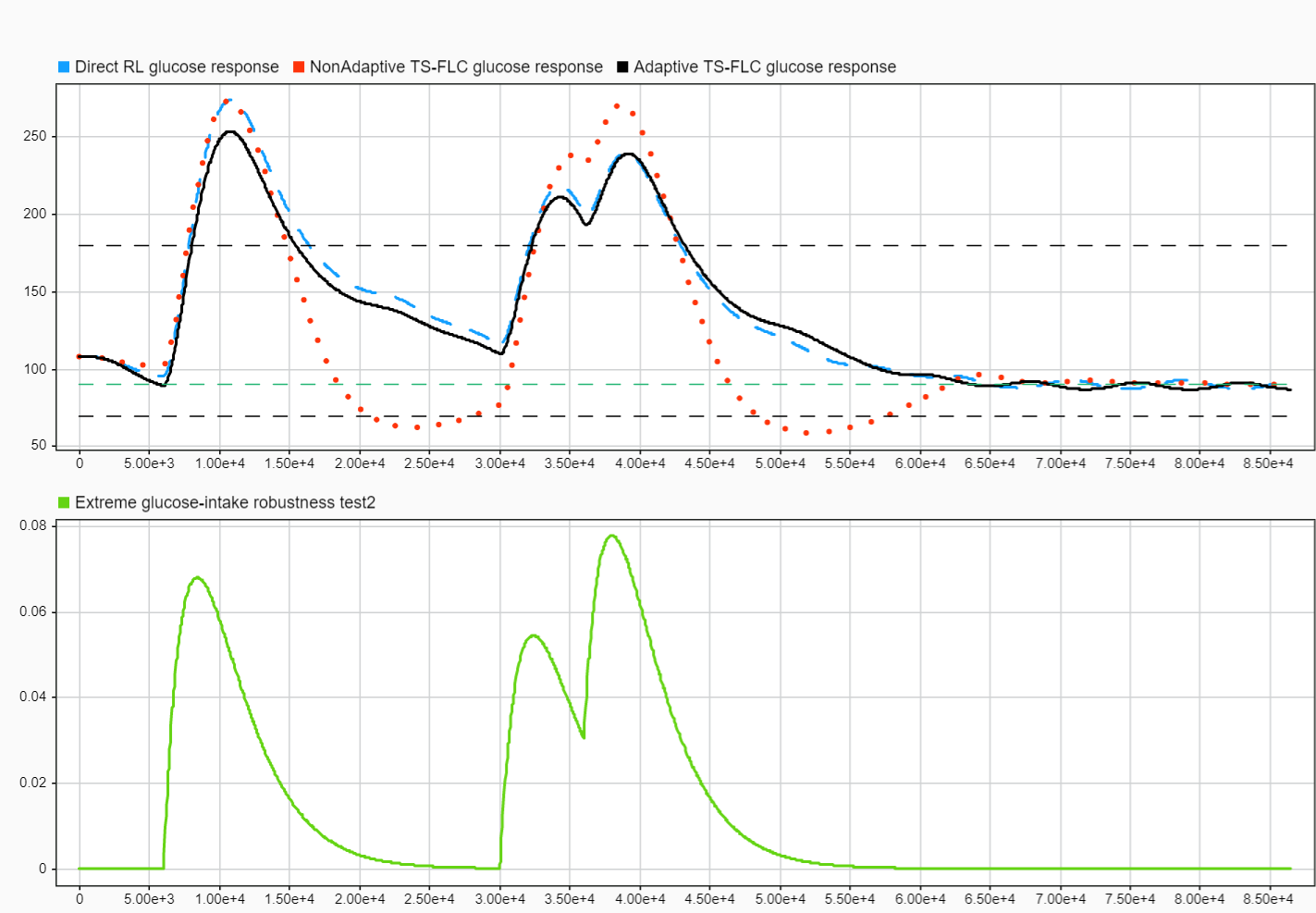}
    \caption{Glucose level (mg/dL) response of the Direct insulin control through Reinforcement Learning, the NonAdaptive Fuzzy Logic Controller and the Adaptive Fuzzy Logic Controller, along with corresponding meals intake (mg/dL).}
    \label{fig:c2}
\end{figure}

We can see from these tests as previous, that when a very large glucose intake is taken, the NonAdaptive fuzzy logic controller, in its attempt to counter the glucose dynamics, undershoots the target value of 90. In contrast, the direct reinforcement learning controller does not undershoot the value 90. Additionally, the proposed Adaptive fuzzy logic controller does not undershoot the value 90 in its effort to counter the glucose dynamics. It also demonstrates a more restrained overshoot of the value 180 compared to both the direct reinforcement learning controller and the non-adaptive fuzzy logic controller.

It's crucial to remember that the tests discussed in this section were conducted under extreme conditions, specifically designed to evaluate the controllers' performance in challenging scenarios.

We should also mention that the average insulin-injection dosage was calculated and ranged between 0.60 and 0.65 units per second.

In summary, the direct RL-based controller demonstrated robust performance in the nominal case, while the non-adaptive fuzzy controller showed promise in terms of stability and robustness analysis. However, the synthesized adaptive fuzzy controller, integrating the best of both worlds, emerged as the most effective solution, offering superior performance in both nominal and non-nominal scenarios. This comparison provides valuable insights for designing controllers in medical applications, emphasizing the importance of adaptability and robustness.

\section{Conclusion}
 Our contributions to artificial pancreas systems, achieved through the integration of reinforcement learning and adaptive fuzzy control, are meticulously outlined. This includes the environment setup, fuzzy controller architecture, and key observations. We synthesized three distinct controllers: direct insulin control through reinforcement learning, non-adaptive fuzzy control, and adaptive fuzzy control. Comparative tests reveal that the adaptive fuzzy controller, which combines reinforcement learning and non-adaptive fuzzy control, demonstrates superior performance in various scenarios. This hybrid approach leverages the adaptability of RL agents and the stability of fuzzy controllers, underscoring the need for continuous optimization of control algorithms to enhance adaptability and robustness.

One significant finding is that reinforcement learning proves to be superior to metaheuristics because it enables adaptive parameter changes, whereas metaheuristics provide static gains. This adaptability is crucial for maintaining optimal performance in the dynamic environment of artificial pancreas systems.

Future research may focus on refining artificial pancreas technology by incorporating additional environmental factors and optimizing control strategies for real-world medical applications. Specifically, future studies could explore other robust architectures, such as PID controllers, and integrate them with the adaptability of reinforcement learning. Moreover, applying fuzzy logic control, optimized through reinforcement learning, can extend to other medical systems like artificial livers and artificial lungs.

\end{document}